\documentclass[12pt,a4paper]{article}
\usepackage[latin1]{inputenc}
\usepackage{amsmath}
\usepackage{mathtools}
\usepackage{setspace}
\usepackage{slashed}
\usepackage{amsfonts}
\usepackage{amssymb}
\usepackage{graphicx}
\usepackage[margin=1in]{geometry}
\usepackage{caption}
\usepackage{subcaption}
\usepackage{tikz}
\usetikzlibrary{shapes.geometric, arrows}
\usepackage[nottoc]{tocbibind}

\usepackage[titletoc]{appendix}
\usepackage{cite}
\usepackage{authblk}
\usepackage{blindtext}
\usepackage{abstract}
\title{\textbf{In-In EFT}}
\author{Namit Mahajan\thanks{ nmahajan@prl.res.in, mahajannamit@gmail.com}}
\affil{Theoretical Physics Division, Physical Research Laboratory, Ahmedabad, India.}
\date{}
\begin{document}
	\maketitle
	\doublespacing
	\begin{abstract}
		The effective field theory (EFT) construction when the objects of interest are the In-In (or real time) correlators rather than the In-Out S-matrix elements is constructed. This is done using the formal equivalence between the In-In and In-Out correlation functions, and demanding that the EFT and the full theory provide the same answers for the In-In correlation functions. Matching coefficients for a simple example are provided including for operators with two or more space or time derivatives. It is also pointed out that the Schrodinger representation of QFT captures some of the desired features of these calcultions in a natural way and may actually serve as a link between these calculations and the broader amplitude programme.
	\end{abstract}
	
	\section{Introduction}
	Effective Field Theories (EFTs) are quite ubiquitous and are routinely utilised (see for example \cite{Weinberg:1978kz}-\cite{Penco:2020kvy}), at times without explicit realisation. The separation of scales leading to effective decoupling of physics at different length/time/energy scales is what actually allows one to make progress and enables to compute physical quantities of interest even in a complicated theory. An example would be computing cross-sections for proton proton scttering. The large center of mass energy involved enables one to express the answer into a convolution of perturbatively calculable partonic cross-section and the parton distribution functions which are non-perturbative inputs (see for example \cite{Collins:1989gx}). It is the separation of physics involved at different scales which allows one to express the cross-section of protons, which are composite objects having a complicated description, in such a way.

	Given a high energy or ultraviolet (UV) theory, the task is to write down a low energy theory (in the Infrared (IR)) such that the two theories make the same predictions for observables of interest when the energies or momenta involved are lower than some reference scale. In its practical avatar, an EFT is written down by identifying all the operators consistent with the symmetries at the low energy scale, built out from the fields that are physical (ie can go on shell) at the relevant scale, and then demanding that the scattering amplitudes computed using the EFT and the full theory are the same at the reference scale. In this version, the quantities of interest are the scattering amplitudes which are nothing but the S-matrix elements computed using the familiar In-Out method. Consider for example a scatteing process in a relativistic theory: $A(p_A) + B(p_B) \rightarrow C(p_c) + D(p_D)$. The initial state comprising of particles (projectiles) $A$ and $B$ with momenta $p_A$ and $p_B$ is prepared in the far past ie $t \to -\infty$. These are free particle states of the given momenta. These then evolve in space and time and come in contact with each other (scatter off or more generally interact). This interaction takes place over a small region in space and over a short interval of time. The final state particles ($C$ and $D$) fly apart and are detected far away from the interaction region in the far future ($t \to \infty$). Thus, in this kind of a situation, the state of the system is known both at the initial and late times. The quantum fields describing the free particle states in the asymptotic regimes ($t \to \pm\infty$) are rlated to the full interacting quantum fields of the theory. This map is provided by the S-matrix (see \cite{Itzykson:1980rh}-\cite{Schwartz:2014sze}). Therefore, the S-matrix elements provide the overlap between the initial (In-state) and the final (Out-state) states. The S-matrix elements are computed from the time ordered Green functions. Following the Lehmann-Symanzik-Zimmermann (LSZ) method, the amplitudes are obtained from these time ordered Green functions by amputating the external legs and supplying the wave functions for the fields,  for example unity for scalars, Dirac spinors for fermions, polarization vectors for photons.

	Contrast the above with the following situation: start with a given state (for simplicity say the vacuum state) in the asymptotic past, $t \to -\infty$. Let the system evolve under a gievn interation Hamiltonian (alternatively interaction Lagrangian). The task at hand now is to compute the expectation value of a given operator (or product of operators) at some later fixed time $t$: $\langle \psi(t)\vert {\mathcal{O}}(t)\vert\psi(t)\rangle$. This is a bona fide physical quantity and such objects are of interest in time dependent situations, like cosmology or when studying non-equilibrium aspects to name a few. The method to deal with such quatities is then aptly called the In-In technique and also goes by the name Schwinger-Keldysh technique or expectation value formalism \cite{Schwinger:1960qe}-\cite{Keldysh:1964ud} (more details and some applications can be found for example in \cite{Chou:1984es}-\cite{Calzetta:2008iqa}). The most important difference compared to the In-Out method is the fact that here the time contour runs from $t = -\infty$ to some finite time $t$, say $t = 0$, and then backwards to $t = -\infty$ to accomplish the task of computing the expectation value in the desired state which started from a given state (for example the vacuum state) at $t = -\infty$. The question of an EFT in such a setting is then vastly different from the more familiar case of matching the amplitudes. However, there is no doubt that an EFT and suitable set of operators commensurate with the symmetries of the theory at the low energy scale must exist. It is the aim of this paper to explore this and show that very similar to the case of scattering amplitudes, one could write down operators built out of 'light fields'. But in contrast to matching the amplitudes on both the sides, we demand the EFT to yield the same In-In correlation functions as the full theory:
	\begin{equation}
	 \langle \textrm{In-In Correlators} \rangle_{\textrm{EFT}} \,\,=\,\, \langle \textrm{In-In Correlators} \rangle_{\textrm{Full\,\,Theory}}
	\end{equation}
This then defines the EFT in the present case - the In-In EFT. As we shall see below, the EFT defined by this equation is not the same as the one obtained by demanding the equality of the amplitudes in both the theories. There are rather very interesting and charecteristic features of the In-In EFT which do not have analogues in the In-Out version. This has to do with the fact that the In-In correlation functions are computed at a specific time. This, thus, breaks time translation invariance, and also the Lorentz invariance, but the spatial rotations are preserved. The spatial momentum conservation will hold, and unlike the scattering amplitudes, only the three dimensional delta function over the spatial momenta will appear rather than the delta function involving four momenta. Therefore, the set of operators that we would end up will differentiate between the spatial and temporal derivatives. In particular, due to the breaking of time translation invariance, the usual practice of throwing away or ignoring the total time derivatives is no longer allowed and this leads to additional surface (boundary) terms. The spatial derivative still can be handled in the usual way, and thus integration by parts involving spatial derivatives proceeds as usual, ignoring the total derivatives.

To keep the discussion concrete, we shall consider two fields in the Minkowski spacetime: a light field $\varphi$ whose mass $m$ is set to zero and a heavy field $\chi$ having mass $M$, interacting via the term $\frac{g}{2}\chi\varphi^2$, with $g$ being a dimensionful coupling constant (it will be useful later on to rewrite $g = M\kappa$, where $\kappa$ is a dimensionless constant). Having $m \neq 0$ doesn't change the essence of the arguments but only complicates some of the intermediate calculations, and the generalization to a non-zero value is straight forward. Since we perform the analysis at the tree level only, this approximation is fine. A one loop analysis would have led to a non-zero value for $m$. Recently, it has been shown that the In-In correlation function and the In-Out correlation function are essentially the same \cite{Donath:2024utn}. Below we rederive this result providing explicit details, and then use it to show how to compute the In-In correlators using the familar In-Out recipe. We then proceed to write down the effective operators consistent with the symmetries of the low energy theory i.e. when the heavy field $\chi$ is no longer present as the dynamical degree of freedom. Using a simple toy Lagrangian for the complete system, we verify that such a construction does indeed provide the right answers. A characteristic feature of such computations is the appearance of the so called total energy pole in the correlation function. This is at unphysical values of the energy (defined more precisely below) and the residue is found to be the scattering amplitude in the high energy limit (see \cite{Arkani-Hamed:2013jha}-\cite{Goodhew:2020hob} for an incomplete list of references, including application to cosmology). We then point out that the Schrodinger representation of QFT, when used to compute the In-Out correlation function or the time ordered Green function for the scattering amplitude, automatically has the total energy pole built into it, and is clearly seen before the limit $t \to \infty$ is taken. This observation, in our opinion, can serve as a bridge between these computations and the general amplitude programme where the natural starting point is the wave function. This perhaps should not be too surprising since the Schrodinger representation of QFT computes the wave-functional of the theory and corrections to it as different interactions are included, completely paralleling the wave-function approach in quantum mechanics.
	
	\section{In-Out vs In-In}
	We begin by recalling some basic and well known facts about the asymptotic free fields and the time evolution operator \cite{Itzykson:1980rh}. The fully interacting field $\varphi$ and the free field at $t = -\infty$, $\varphi_{In}$, are simply related by
	\begin{equation}
	 \lim_{t \to -\infty} \varphi(t,\vec{x}) = \sqrt{Z}\, \varphi_{In}(t,\vec{x})
	\end{equation}
where $Z$ is the wavefunction renormalization constant. Similar equation holds for $\varphi_{Out}$ at $t = \infty$.
Expressed in terms of the time evolution operator, $U(t)$, the interacting field at any given time $t$ is expressed as
\begin{equation}
 \varphi(t,\vec{x}) = U^\dag(t)\,\varphi_{In}(t,\vec{x})\, U(t)
\end{equation}
where $U^\dag = U^{-1}$ is implicit here. Also, in view of the above, $\lim_{t \to -\infty}U(t) = 1$. Further,
\begin{equation}
 U(t,t') = U(t) U^{-1}(t')\,\,\,\, , \,\,\,\, U(t) = U(t,t')\,U(t') \,\,\,\, \textrm{and}\,\,\,\, U(t_1,t_2)\,U(t_2,t_3) = U(t_1,t_3)
\end{equation}
such that the S-matrix is defined as
\begin{equation}
 \lim_{t \to \infty} \lim_{t' \to -\infty} \,U(t,t') = S
\end{equation}
It is convenient, and useful, to exprss $U(t)$ in the following form:
\begin{equation}
 U(t) = T\left(\exp\left[-i\int_{-\infty}^t\,dt'\,H_{int}(t')\right]\right)\,\underbrace{U(-\infty)}_{1\!\!1} = T\left(\exp\left[-i\int_{-\infty}^t\,dt'\,H_{int}(t')\right]\right)
\end{equation}
Therefore, one immediately arrives at
\begin{equation}
 S = \lim_{t_f \to \infty} \lim_{t_i \to -\infty} \underbrace{T\left(\exp\left[-i\int_{-t_i}^{t_f}\,dt\,H_{int}(t)\right]\right)}_{U(t_f,t_i)} = T\left(\exp\left[i\int\,d^4x\,{\mathcal{L}}_{int}(x)\right]\right)
\end{equation}
where in the last step we have assumed that there are no derivative interactions. In terms of the states,
\begin{equation}
 S\vert \beta_{Out}\rangle = \vert\beta_{In}\rangle\,\,\, \textrm{such that}\,\,\,\, S_{\beta\alpha} = \langle\beta_{Out}\vert\alpha_{In}\rangle
\end{equation}
Assuming the vacuum state to be stable, the action of $S$ on vacuum leads to an unmeasurable phase: $S\vert 0\rangle = e^{i\theta}\vert 0\rangle$.

Next let us consider the definitions of the In-In and In-Out correlation functions: $G_{\textrm{In-In(In-Out)}}(x_1,x_2...x_n)$. Consider preparing a quantum system in free theory vacuum state at an early time: $t = -\infty$. The system then evolves according to the interaction Hamiltonian. The expectation value that we are interested in, $\langle {\mathcal{O}}(t)\rangle$ is
\begin{equation}
 \langle {\mathcal{O}}(t)\rangle = \langle 0\vert U(-\infty,t) {\mathcal{O}} U(t,-\infty)\vert 0\rangle = \langle 0\vert U(-\infty,t) \underbrace{U(t,\infty)U(\infty,t)}_{= {1\!\!1}} {\mathcal{O}} U(t,-\infty)\vert 0\rangle
\end{equation}
This can be generalised to the case when the operator ${\mathcal{O}}$ is a product of field operators (at same or different times). The insertion of identity in the above equation was on the so called forward branch, but could have been on the backward branch as well. Now, combine the two left most $U$ operators, $U(-\infty,t) U(t,\infty) = U(-\infty,\infty)$, and express everything in terms of the free field operators $\varphi_{In}$. Next, writing $U(-\infty,\infty) = \left(T\,\exp\left[-i\int_{-\infty}^{\infty}\,dt\,H_{int}(t)\right] \right)^{\dag} = \bar{T}\,\left(\,\exp\left[i\int_{-\infty}^{\infty}\,dt\,H_{int}(t)\right] \right) (= U^{\dag}(\infty,-\infty) = U^{-1}(\infty,-\infty))$, the In-In correlation function can be written as:
\begin{equation}
 G_{\textrm{In-In}}(x_1,...x_n) = \langle 0\vert \bar{T}\left(\,\exp\left[i\int_{-\infty}^{\infty}\,dt\,H_{int}(t)\right] \right) T\left(\prod_i \varphi_{in}(x_i) \,\exp\left[-i\int_{-\infty}^{\infty}\,dt\,H_{int}(t)\right]\right)\vert 0\rangle
\end{equation}
It is to be remembered that the upper limits in the above equation need not be $+\infty$ but any time value greater than the largest of the $t_i$. It is easy to verify that any time evolution beyond this largest value cancels between the $T$ and $\bar{T}$ parts in the above equation. This is the reason why in the In-In correlation functions, instead of $+\infty$ often some finite value say $\bar{t}$ appears (often the choice $\bar{t} = 0$ is made).

We now proceed to the In-Out correlation function or the time ordered Green function initially written in terms of the interacting fields:
\begin{equation}
 G_{\textrm{In-Out}}(x_1,x_2...x_n) = \langle 0\vert T[\varphi(x_1)...\varphi(x_n)]\vert 0\rangle
\end{equation}
Express in terms of the In fields: $\varphi = U^{-1}\varphi_{In}U$. Consider $t_1>t_2>....>t_n$ and make use of the following: $U(t_i)U^{-1}(t_j) = U(t_i,t_j)$, $U^{-1}(t_1) = U^{-1}(t)U(t,t_1)$ and $U(t_n) = U(t_n,-t)U(-t)$ where we have chosen some arbitrary time $t$ such that $t >> t_1$ and $-t << t_n$. We thus have:
\begin{equation}
 G_{\textrm{In-Out}}(x_1,x_2...x_n) = \langle 0\vert U^{-1}(t) T[\prod_i \varphi_{In}(x_i)\, \exp\left[-i\int_{-t}^t dt' H_{int}(t')\right]] U(-t)\vert 0\rangle
\end{equation}

And as appropriate for scattering amplitudes, $t\to \infty$. Then the action of $U(-\infty)$ on $\vert 0\rangle$ and that of $U^{-1}(\infty)$ on $\langle 0\vert$ leads to the normalization factor $(\langle 0\vert S\vert 0\rangle)^{-1} = (\langle 0\vert T\left(\exp\left[-i\int_{-\infty}^{\infty}\,dt\,H_{int}(t)\right]\right) \vert 0\rangle)^{-1}$. This is what leads to the vacuum bubbles, and the time ordered Green function takes the familiar form
\begin{equation}
 G_{\textrm{In-Out}}(x_1,x_2...x_n) = \frac{\langle 0\vert T[\prod_i \varphi_{In}(x_i)\, \exp\left[-i\int_{-t}^t dt' H_{int}(t')\right]]\vert 0\rangle}{\langle 0\vert T\left(\exp\left[-i\int_{-\infty}^{\infty}\,dt\,H_{int}(t)\right]\right) \vert 0\rangle}
\end{equation}
Instead of proceeding like this, if one makes use of Eq.(6) in Eq.(12), with $t\to \infty$, one would immediately arrive at
\begin{eqnarray}
 G_{\textrm{In-Out}}(x_1,x_2...x_n) &=& \langle 0\vert \bar{T}\left(\,\exp\left[i\int_{-\infty}^{\infty}\,dt\,H_{int}(t)\right] \right)T\left(\prod_i \varphi_{In}(x_i)\, \exp\left[-i\int_{-\infty}^{\infty} dt' H_{int}(t')\right]\right) \vert 0\rangle \nonumber \\
 &=& G_{\textrm{In-In}}(x_1,...x_n)
\end{eqnarray}
After having explicitly rederived the result about the In-In correlation function being the same as the In-Out one, we show a quick calculation where the In-In correlator is calculated using the In-Out methods and the usual Feynman rules, closely following \cite{Donath:2024utn}. In what follows, we take the time argument of all the external fields to be the same as is often the case say while computing cosmological correlators of interest. Further, recall that given an interaction Hamiltonian or Lagrangian, and using the Feynman rules derived from it, it is straightforward to write down an In-Out correlator or time ordered Green function. It is important to note that one needs to implement the correct $'i\epsilon'$ prescription to pick out the right contribution from the poles. In the In-Out case, we use the prescription: $m^2 \to m^2 + i\epsilon$, which can be explicitly derived using the integral representation of the $\Theta$ function or can be motivated from the argument of the convergence of the Path Integral (PI). Also, recall that in the mometum space, the contour is picked such that the pole in the complex $p^0$ plane which is on the negative axis is contained within the contour. With this in mind, the In-In correlation function can be computed following the steps below. For concretness, consider the In-In correlation function of four light fields interacting via the interaction Lagrangian described earlier:
$\frac{g}{2}\chi\varphi^2$, where we assume that $m_{\varphi}=0$ and $m_{\chi} = M$. There are three different contributions which are usually referred to as $s$, $t$ and $u$-channel contributions or diagrams in analogy to the usual scattering amplitudes. It is to be remembered that this is just an analogy and the fact that only three momenta are conserved. With all the momenta as incoming, we have
\begin{equation}
 G_{\textrm{In-Out}}(x_1,x_2...x_n) = \int \prod_j \frac{d^4p_j}{(2\pi)^4} \,\exp\left(i\sum_j p_j\cdot x_j\right) \underbrace{(2\pi)^4 \delta^4(\sum_j p_j) G_{\textrm{In-Out}}(p_1,p_2...p_n)}_{\tilde{G}_{\textrm{In-Out}}(p_1,p_2...p_n)}
\end{equation}
The equal time Fourier transform of $\tilde{G}_{\textrm{In-Out}}(p_i)$ is then the real time In-In correlation function:
\begin{equation}
G_{\textrm{In-In}}(t) = (2\pi)^3 \delta^3(\sum_j \vec{p}_j) \int \prod_j \frac{d\omega_j}{(2\pi)}\,(2\pi) \delta(\sum_j \omega_j)\,\exp(i\,t\sum_j \omega_j)\,\tilde{G}_{\textrm{In-Out}}(p_i)
\end{equation}
The overall factor $(2\pi)^3 \delta^3(\sum_j \vec{p}_j)$ is due to the three momentum conservation and will not be explicitly shown in what follows below. For the case at hand, for the $s$-channel contribution, we have
\begin{equation}
 {G}^{(s)}_{\textrm{In-Out}}(p_1,p_2,p_3,p_4) = (-ig)^2 \frac{i}{(p_1+p_2)^2-M^2 + i\epsilon} \prod_{i=1}^4\frac{i}{(p_i^2 + i\epsilon)}
\end{equation}
Defining the followng quantities: $E_i^2 = \vert\vec{p}_i\vert^2$, $E_{ij} = E_i+E_j$, $E_{s}^2 = \vert\vec{p}_1+\vec{p}_2\vert^2 + M^2 = \vert\vec{p}_3+\vec{p}_4\vert^2 + M^2$, $\omega_{s} = \omega_1 + \omega_2 = \omega_3 + \omega_4$, $p_{ij}^2 = \vert \vec{p}_i+\vec{p}_j\vert^2 = E_i^2 + E_j^2 + 2\vec{p}_i\cdot\vec{p}_j$, and the fact that three momentum conservation implies $\vec{p}_1+\vec{p}_2 = -(\vec{p}_3+\vec{p}_4)$, thus implying $p_{12}^2 = p_{34}^2$,
\begin{equation}
 {G}^{(s)}_{\textrm{In-Out}}(p_1,p_2,p_3,p_4) = (-ig)^2 \frac{i}{\omega_{s}^2-E_{s}^2 + i\epsilon} \prod_{i=1}^4\frac{i}{(\omega_i^2 - E_i^2 + i\epsilon)}
\end{equation}
Thus,
\begin{equation}
 G^s_{\textrm{In-In}}(t;\vert\vec{p}_i\vert) =  -ig^2\int_{-\infty}^{\infty} \prod_j \frac{d\omega_j}{(2\pi)}\,\frac{(2\pi)\, \delta(\sum_j \omega_j)\,\exp(i\,t\sum_j \omega_j)}{(\omega_{12}^2-E_{12}^2 + i\epsilon) \prod_{i=1}^4(\omega_i^2 - E_i^2 + i\epsilon)}
\end{equation}
It is now straightforward to compute the residues at poles which lie on the negative axis for each of the $\omega_i$. One finally obtains, with $E_T = \sum_{i=1}^4E_i$
\begin{equation}
 G^s_{\textrm{In-In}}(t;\vert\vec{p}_i\vert) =  -i\frac{g^2}{8E_1E_2E_3E_4E_T}\, \left(\frac{(E_s+E_T)}{E_s(E_s+E_{12})(E_s+E_{34})}\right)
 \label{Ginin-rep}
\end{equation}
This result agrees with the direct calculation employing the Schwinger-Keldysh approach. The pole at $E_T = 0$, the total energy pole, is now clear. This is at unphysical values. In a scattering or decay calculation, the pole would signify a $2 \to 2$ or, in this case, $1 \to 3$ process. Hoever, in this case, it is like $0 \to 4$ process. Clearly, this does not happen in vacuum and is not a prototypical S-matrix element.

The term in the parenthesis can now be expanded in inverse powers of $M$ after expressing different quantities in a conveneint, symmetric form like $E_s = \sqrt{\frac{p_12^2}{2}+ \frac{p_34^2}{2}+ M^2}$. First few terms in such an expansion are:

{\bf 1.} $\mathbf{\mathcal{O}}(\frac{1}{M^2})$: $1$

{\bf 2.} $\mathbf{\mathcal{O}}(\frac{1}{M^3})$: $E_T - E_{12} - E_{34} = 0$

{\bf 3.} $\mathbf{\mathcal{O}}(\frac{1}{M^4})$: $\frac{1}{2} \left[-p_{12}^2 - p_{34}^2 - 2(E_1+E_2)(E_3+E_4)\right]$

{\bf 4.} $\mathbf{\mathcal{O}}(\frac{1}{M^5})$: $E_T(E_1+E_2)(E_3+E_4)$

As is clear from the form of the terms shown above, there is a symmetry as expected for the $s$-channel case. Similar expressions can be worked out for the $t$ and $u$-channels as well, and can be read off directly using the crossing symmetry. Such an expansion in terms of inverse powers of $M$ suggests an EFT construction. Compared to the more familair EFT for scattering amplitudes, there are terms with odd inverse powers on $M$, like the non-zero term ${\mathcal{O}}(\frac{1}{M^5})$ above. Such terms, as we shall see below, stem from the fact that the time translation invariance is broken leading to operators with time derivatives which would have been ignored otherwise.

\section{Effective Field Theory}
We follow the general principle of constructing an EFT: write down all opertors to a given dimension commensurate with the symmetries of the theory at low energy. In the present case, the breaking of time translations means that the spatial and temporal derivatives are no longer at the same footing. We still have $\varphi \to -\varphi$ symmetry, thereby forbidding odd powers of $\varphi$ in the effective Lagrangian. As an example, consider the operator $\partial_{\mu}(\varphi^2\partial^{\mu}\varphi^2)$. This can be written as
\begin{equation}
 \partial_{\mu}(\varphi^2\partial^{\mu}\varphi^2)= \partial_{t}(\varphi^2\partial_{t}\varphi^2) - \partial_{i}(\varphi^2\partial_{i}\varphi^2) = \partial_{\mu}\varphi^2\partial^{\mu}\varphi^2 + \varphi^2\Box\varphi^2
\end{equation}
Normally, the left hand side would have been set to zero, it being a total derivative, and thus, the two operators on the right hand side are not independent and are equivalent. Choosing one over the other defines the operator basis. When time translations are broken, but spatial translations are preserved, the spatial total derivative can still be set to zero but the temporal derivative term, $\partial_{t}(\varphi^2\partial_{t}\varphi^2)$ survives and will yield a surface or boundary term. This operator, then, is an independent additional one in the EFT description. Since we are here interested in the In-In correlation function involving four light fields, we write down operators with four fields and arbitrary number of derivatives, spatial and temporal and their combinations obeying the symmetries - spatial rotations in this case.

{\bf Operators with Spatial Derivatives}: These operators have the spatial derivatives contracted appropriately. At the two derivative order, we have:
\begin{equation}
 (\partial_{i}\varphi)(\partial_{i}\varphi)\varphi^2, \,\,\,\,(\partial_{i}\varphi^2)(\partial_{i}\varphi^2)\,\,\,\, \textrm{and}\,\,\, \varphi^2(\partial_{i}^2\varphi^2)
\end{equation}
It is straightforward to convince onself that out of these three, one could choose to work with  $\varphi^2(\partial_{i}^2\varphi^2)$ as the independent one. Let us look at this term in some detail:
\begin{eqnarray}
 \varphi^2(\partial_{i}^2\varphi^2) &\to& \varphi_a\varphi_b(\partial_{i}^2\varphi_c\varphi_d) = \varphi_a\varphi_b \partial_i[\partial_i(\varphi_c\varphi_d)] \nonumber \\
 &\rightarrow& -p_{12}^2\,\varphi^4 + \textrm{different permutations}
\end{eqnarray}
where $\varphi_a$ is a shorthand used to indicate $\varphi(p_a)$. We therfore note that $\partial_i^2$ yields the combination $p_{ab}^2 = \vert \vec{p}_a+\vec{p}_b\vert^2$, where $\vec{p}_{a,b}$ are the three momenta of the two fields on which it acts. Higher spatial derivatives can be manipulated similarly. In the above expression, $\varphi^4$ has been explicitly written after the action of the derivatives. This is to act as a reminder that unlike the case of dealing with scattering amplitudes, the propagators or the two point functions for the external fields are to be written and not amputated. It is these external propagators that lead to the $E_1E_2E_3E_4$ factor in the denominator of Eq.(\ref{Ginin-rep}).

{\bf Operators with Time Derivatives}: Due to the breaking of time translations, odd powers of $\partial_t$ can also appear:

1. $\partial_t\varphi^2$: This gives $(\omega_a+\omega_b)\,\varphi^2$

2. $\partial_t\varphi^4$: This gives $(\omega_a+\omega_b+\omega_c+\omega_d)\,\varphi^4$

3. $\partial_t^2\varphi^2$: Leads to $(\omega_a+\omega_b)^2\,\varphi^2$

4. $\partial_t^2\varphi^4$: This gives $(\omega_a+\omega_b+\omega_c+\omega_d)^2\,\varphi^4$

5. $\partial_t\varphi^2\partial_t\varphi^2$: This operator yields $(\omega_a+\omega_b)(\omega_c+\omega_d)\,\varphi^4$

6. $(\partial_t^2\varphi^2)(\partial_t\varphi^2)$: This operator or $\partial_t(\partial_t\varphi^2 \partial_t\varphi^2)$ which is clearly a surface term and usually set to zero is an important one. This leads to $(\omega_a+\omega_b)^2(\omega_c+\omega_d)\,\varphi^4$ and permutations.

{\bf Operators with Mixed Spatial and Time Derivatives}: The lowest among them has one time and two space derivatives:

1. $\partial_t(\varphi^2 \partial_i^2\varphi^2)$ which yields terms like $(\omega_a+\omega_b+\omega_c+\omega_d)\,p_{cd}^2\,\varphi^4$ and permutations.

2. $\partial_t\varphi^2 \partial_i^2\varphi^2$: yields terms like $(\omega_a+\omega_b)\,p_{cd}^2\,\varphi^4$ and permutations.

Operators with increasing mass dimension can be systematically written in an analogous fashion. It may be worthwhile to recall that since the main interest here is to consider the In-In correlator of four light fields, the discussion of the higher dimensional operators has been restricted to the case of operators with four fields and derivatives. In reality, say at dimension six, there will be operators like $\varphi^6$ also present. These will play a very important role when loop effects are considered. For the time being, we proceed with the restricted set only. We thus have the effective description suited for In-In correlation function calculations described by the following low energy Lagrangian:
\begin{eqnarray}
 {\mathcal{L}}_{\textrm{EFT}} &=& \frac{1}{2}(\partial_t\varphi)^2 - \frac{1}{2}v^2 (\partial_i\varphi)^2 - m^2\varphi^2 + C_3 \partial_t\varphi^2 - \frac{\lambda}{4!}\varphi^4 + C_4 \partial_t^2\varphi^2 + C_5\partial_t\varphi^4 + C_{6,1}\varphi^2(\partial_{i}^2\varphi^2) \nonumber \\
 && + C_{6,2}\partial_t^2\varphi^4 + C_{6,3}\partial_t\varphi^2\partial_t\varphi^2 + C_{6,4}\varphi^6 +
C_{6,5}(\partial_{i}^2)^2\varphi^2 + C_{6,6}\partial_t^4\varphi^2 \nonumber \\
&& +  C_{7,1}\partial_t(\varphi^2 \partial_t^2\varphi^2) + C_{7,2}\partial_t\varphi^2 \partial_i^2\varphi^2 + .....
\end{eqnarray}
where due to the boundary term being non-zero in the present case, we've distinguished between $(\partial_t\varphi)^2$ and $\partial_t^2\varphi^2$, and to highlight that the spatial and temporal derivatives are no longer at the same footing, we have explicitly introduced a factor of velocity $v$ along with the $(\partial_i\varphi)^2 $ term. Also, the coefficients $C_{6,5}$ and $C_{6,6}$ need not be the same (ie simply different upto a trivial sign factor) since we do not attempt to combin the two terms and express as $\Box^2$ plus additional terms. Some of these aspects may not be visible in a tree level matching calculation, say $v$ being different than unity (velocity of light $c=1$ in natural units). However, it is easy to convince oneself that once loop effects are included, these features will show up.

Having written down a set of effective operators, we next compute their contribution to the four point In-In correlation function. Working at the tree level, we use the method discussed in the previous section to compute these contributions and then match with the toy example considered above to determin the Wilson coefficients. Below, we denote different contribution by $G^{\textrm{In-In}}_4(\textrm{coupling})$.

1. $\frac{\lambda}{4!}\varphi^4$:

\begin{eqnarray}
 G^{\textrm{In-In}}_4(\lambda) &=& -{i\lambda}\int_{-\infty}^{\infty} \prod_j \frac{d\omega_j}{(2\pi)}\,\frac{(2\pi)\, \delta(\sum_j \omega_j)\,\exp(i\,t\sum_j \omega_j)}{ \prod_{i=1}^4(\omega_i^2 - E_i^2 + i\epsilon)} \nonumber \\
 &=& -i\lambda \frac{1}{8E_1E_2E_3E_4E_T}
\end{eqnarray}
Comparing with the appropriate term (in this case the constant term in the expansion), one finds (and recall that $g=\kappa M$
\begin{equation}
 \lambda  =  \frac{g^2}{M^2} = \kappa^2
\end{equation}

2. $C_{7,1}\partial_t(\varphi^2 \partial_i^2\varphi^2)$:
\begin{eqnarray}
 G^{\textrm{In-In}}_4(C_{7,1}) &=& iC_{7,1}\int_{-\infty}^{\infty} \prod_j \frac{d\omega_j}{(2\pi)}\,\frac{(2\pi)\, \delta(\sum_j \omega_j)\,\exp(i\,t\sum_j \omega_j)}{ \prod_{i=1}^4(\omega_i^2 - E_i^2 + i\epsilon)} (\omega_1+\omega_2+\omega_3+\omega_4)[p_{12}^2+ \textrm{perm.}]\nonumber \\
 &=& 0
\end{eqnarray}
The presence of $\delta(\sum_j \omega_j)$ leads to a vanishing result, independent of $C_{7,1}$. This operator doesn't make its presence felt to this particulator correlator.

3. $C_{6,1}\varphi^2(\partial_{i}^2\varphi^2)$:
\begin{eqnarray}
 G^{\textrm{In-In}}_4(C_{6,1}) &=& -iC_{6,1}\int_{-\infty}^{\infty} \prod_j \frac{d\omega_j}{(2\pi)}\,\frac{(2\pi)\, \delta(\sum_j \omega_j)\,\exp(i\,t\sum_j \omega_j)}{ \prod_{i=1}^4(\omega_i^2 - E_i^2 + i\epsilon)} (p_{12}^2+p_{34}^2 + t + u)\nonumber \\
 &=& iC_{6,1}\frac{(p_{12}^2+p_{34}^2 + t + u)}{8E_1E_2E_3E_4E_T}
\end{eqnarray}

4. $C_{6,3}\partial_t\varphi^2\partial_t\varphi^2$: This involvs a bit of manipulation. The time derivatives yield $(\omega_1+\omega_2)(\omega_3+\omega_4)$ which upon use of $\delta(\sum_j \omega_j)$ can be written as $-(\omega_1+\omega_2)^2$. Intermediate steps in carrying out the successive integrals get a bit messier but in the end the algebra simplies.
\begin{eqnarray}
 G^{\textrm{In-In}}_4(C_{6,3}) &=& -iC_{6,3}\int_{-\infty}^{\infty} \prod_j \frac{d\omega_j}{(2\pi)}\,\frac{(2\pi)\, \delta(\sum_j \omega_j)\,\exp(i\,t\sum_j \omega_j)}{ \prod_{i=1}^4(\omega_i^2 - E_i^2 + i\epsilon)} (\omega_1+\omega_2)^2\nonumber \\
 &=& iC_{6,3}\frac{(E_1+E_2)(E_3+E_4)}{8E_1E_2E_3E_4E_T}
\end{eqnarray}

The last two contributions can be compared with the ${\mathcal{O}}(1/M^4)$ term in the expansion to extract $C_{6,1}$ and $C_{6,3}$.

5. $C_{7,1}\partial_t(\varphi^2 \partial_t^2\varphi^2)$: As the last example, we compute the contribution due to this operator. Again, like the previous calculation, use of $\delta(\sum_j \omega_j)$ is made to obtain $-(\omega_1+\omega_2)^3$. Rest of the calculation proceeds in the same fashion.
\begin{eqnarray}
 G^{\textrm{In-In}}_4(C_{7,1}) &=& -C_{7,1}\int_{-\infty}^{\infty} \prod_j \frac{d\omega_j}{(2\pi)}\,\frac{(2\pi)\, \delta(\sum_j \omega_j)\,\exp(i\,t\sum_j \omega_j)}{ \prod_{i=1}^4(\omega_i^2 - E_i^2 + i\epsilon)} (\omega_1+\omega_2)^3\nonumber \\
 &=& C_{7,1}\frac{(E_1+E_2)(E_3+E_4)E_T}{8E_1E_2E_3E_4E_T}
\end{eqnarray}
Immediate comparison with the ${\mathcal{O}}(1/M^5)$ term in the expansion above yields
\begin{equation}
 C_{7,1} = -i\frac{g^2}{M^5} = -i\frac{\kappa^2}{M^3}
\end{equation}
The purely imaginary Wilson coeficient is due to the presence of three derivatives.

We have thus demonstrated that there is a simple and efficient way to write down the effective operators and thus defining the In-In EFT Lagrangian. The characteristic features are the presence of odd number of derivatives and the fact that the coefficients of $\partial_t^2$ and $\partial_i^2$ ar not the same in general. This is due to the breaking of time translations, and thus also Lorentz invariance. The presence of odd number of derivatives, or terms with odd inverse powers of $M$, in a purely scalar theory does not happen when constructing effective theory for scattering amplitudes. However, the fact that now odd number of derivatives are allowed, no longer makes it surprising or unexpected.

Having written down the effective operators then gives us immediate tools to compute the desired correlation functions, including the corrections - both loop as well as higher derivative. This approach is better suited to compare with the experiments since having a 'complete model' in mind and then trying to identify the effective operators, only a subset of all possible operators at a given dimension appears corresponding to the complete theory. But some operators at the same dimension will not be there. An example of this is $\partial_t(\varphi^2\partial_t\varphi^2)$ in the above example. This would lead to differences in the Wilson coefficients compared to a direct comparison with the toy model we had in mind as a complete theory. But this would have contributed differently to the correlation function than  $\partial_t\varphi^2\partial_t\varphi^2$. The drawback of being model independent is that there would be too many operators with coefficients which have to be taken to be independent. Still, for problems in cosmology or non-equilibrium studies, such an approach may be more beneficial.\\

{\bf Comparison with the literature:}
To the best of our knowledge, this is the first attempt to write down an EFT for the In-In correlatos in this direct fashion. It is noteworthy to mention the following two references that construct effective operators, including the ones involving time derivatives. Calculations for In-In correlation functions for light fields are performed assuming a model Lagrangian or Hamiltonian describing the interactions of light fields with some heavy fields. A large mass expansion is then performed and then effective operators are identified. These studies necesarily need to identify the right boundary terms and manipulate these to a suitable form:\\
(i) \cite{Salcedo:2022aal} is devoted to the analytic properties of the wavefunction and as a concrete example and application of a UV complet theory consider two scalar fields interacting with each other. Using the equations of motion, but now with the homogeneous part of the solution as well while expressing the heavy field in terms of the light field (as is often done as a recipe for integrating out), the authors arrive at a low energy theory with higher derivatives;\\ (ii) \cite{Green:2024cmx} which contains a direct and brute force calculation of the In-In correlators as well as gets to the same results via the wavefunction approach and arguments based on Wilsonian RG. The authors again make use of the equations of motion and a lot of care is needed because of the presence of explicit temporal derivative terms as well.

 The approach presented here also takes into account the boundary terms but is rather independent of the specific interaction(s) assumed or adopted. A quick comparison then reveals complete agreement while at the same time showing that the method presented here is more general and can easily acommodate a wider set of interactions at the effective level and relies directly on the symmetries of the low energy theory, now suitable for In-In calculations, in a manner similar to the usual EFTs that we are more familiar with.

\section{Schrodinger Representation of QFT - A Short Detour}
Schrodinger representation or picture (see for example \cite{Symanzik:1981wd}-\cite{Hatfield:2019sox})is not very commonly used. It is similar in spirit to the Schrodinger wave function method in quantum mechanics. We provide a brief review, first considering the free theory. More details can be found in \cite{Hatfield:2019sox}. Given a quantum field $\varphi(\vec{x})$ which is time independent, let $\vert\phi\rangle$ be an eigenstate: $\varphi(\vec{x})\vert\phi\rangle = \phi(\vec{x})\vert\phi\rangle$. Analogous to the wave function and it's coordinate reprsentation in ordinary quantum mechanics, the coordinate representation of the quantum state of the given system, $\vert\Psi\rangle$ is the wave-functional, $\Psi[\phi] = \langle\phi\vert\Psi\rangle$. Similar to quantum mechanics, one writes the conjugate momentum operator in terms of functional derivative of $\phi$ and it's matrix elements:
\begin{equation}
 \pi(\vec{x}) =-i\frac{\delta}{\delta\phi(\vec{x})}, \,\,\,\,\,\,\,\, \langle\phi'\vert\pi(\vec{x})\vert\phi\rangle = -i\frac{\delta}{\delta\phi(\vec{x})}\delta[\phi-\phi']
\end{equation}
The Schrodinger equation, $i\frac{\partial}{\partial t}\vert\Psi\rangle = H\,\vert\Psi\rangle$ is cast into a functional differential equation
\begin{equation}
 i\frac{\partial}{\partial t}\Psi[\phi,t] = \frac{1}{2}\int d^3x \left(-\frac{\delta^2}{\delta\phi^2(\vec{x})} + \vert\nabla\phi\vert^2+ m^2\phi^2\right)\Psi[\phi,t]
\end{equation}
For a time independent Hamiltonian, one can then write $\Psi[\phi,t] = e^{-iEt}\Psi[\phi]$, where $\Psi[\phi]$ satisfies the time independent functional Schrodinger equation. Proceeding similar to quantum mechanics, with derivatives replaced by functional derivatives etc, and on the way fixing the overall normalization, one arrives at the wavefunctional in the Fourier space ($\tilde{\phi}(\vec{k})$, the Fourier transform of $\phi(\vec{x}))$. The free theory wavefunctional in the $\tilde{\phi}$ space then takes the form (with $\omega_k = \sqrt{\vec{k}^2+m^2})$
\begin{equation}
 \Psi_0[\tilde{\phi}] = \prod_{\vec{k}}\left(\frac{\omega_k}{\pi}\right)^{\frac{1}{4}}\,\exp\left(-\frac{1}{2}\frac{1}{(2\pi)^3}\omega_k\,\tilde{\phi}^2(\vert\vec{k}\vert)\right)
\end{equation}
This is nothing but the equivalent of free harmonic oscillator wave function. The creation and annihilation operators can be analogously expressed in terms of $\phi$ and $\pi$:
\begin{eqnarray}
 a(\vec{k}) = \int d^3x e^{i\vec{k}\cdot\vec{x}}\left(\omega_k\phi(\vec{x}) + \frac{\delta}{\delta\phi(\vec{x})}\right) \nonumber\\
  a^{\dag}(\vec{k}) = \int d^3x e^{-i\vec{k}\cdot\vec{x}}\left(\omega_k\phi(\vec{x}) - \frac{\delta}{\delta\phi(\vec{x})}\right)
\end{eqnarray}
One can easily verify that $a(\vec{k})\Psi_0[{\phi}] = 0$ and that the first excited state $\Psi_1[{\phi}] = \frac{1}{\sqrt{(2\pi)^32\omega_{k_1}}}a^{\dag}(\vec{k})\Psi_0[{\phi}]$, and so on, completely analogous to harmonic oscillator in quantum mechanics.

Since in the functional Schrodinger representation, the dynamics is encoded in the states, one simply computes the states in the interacting, say, via the perturbation theory for weak couplings and then takes an overlap between the initail and final states. This will immediately yield the S-matrix. To the free Hamiltonia, add the interaction term $\frac{\lambda}{4!}\phi^4(\vec{x})$. The n-th wavefunctional and the energy eigenvalue are then written in terms of the free theory quantities:
\begin{eqnarray}
\Psi_n[{\phi}] =  \Psi_n^{(0)}[{\phi}] + \lambda \Psi_n^{(1)}[{\phi}] + .....\nonumber \\
E_n = E_n^{(0)} + \lambda E_n^{(1)} + ....
\end{eqnarray}
where $H_0\Psi_n^{(0)}[{\phi}] = E_n^{(0)}\Psi_n^{(0)}[{\phi}]$, where the Hamiltonian is decomposed into the free part and the interacting part: $H = H_0 + \lambda H_{int}$. The next steps are exactly the same as in quantum mechanics.

The S-matrix (or the time ordered Green function) for four external fields can be writtn as:
\begin{eqnarray}
 S_{\beta\alpha} &=& \int {\mathcal{D}}\phi\,\, \Psi^*_{p_3p_4}[\phi],\Psi_{p_1p_2}[\phi] \nonumber\\
 &=& \int {\mathcal{D}}\phi\,\,\Psi^*_{0}[\phi]\,a(\vec{p}_3)\,a(\vec{p}_4)\,a^{\dag}(\vec{p}_1)\,a^{\dag}(\vec{p}_2)\,\Psi_{0}[\phi] \\
 &=& \int {\mathcal{D}}\phi\,\, \left(\Psi^{*(0)}_{0}[\phi] + \Psi^{*(1)}_{0}[\phi] +...\right)\,a(\vec{p}_3)\,a(\vec{p}_4)\,a^{\dag}(\vec{p}_1)\,a^{\dag}(\vec{p}_2)\,\left(\Psi^{(0)}_{0}[\phi] + \Psi^{(1)}_{0}[\phi] +...\right) \nonumber
\end{eqnarray}
Skipping the intermediate algebra and straightforward manipulations, to ${\mathcal{O}}(\lambda)$, we arrive at the following form of the S-matrix after switching on the trivial time dependence:
\begin{eqnarray}
 S_{\beta\alpha} &\sim& -\frac{\lambda}{4!}\int\prod_{j=1}^4\left(\frac{d^3\vec{k}_j}{(2\pi)^3} \right) (2\pi)^3\delta^3(\vec{k}_1+\vec{k}_2+\vec{k}_3+\vec{k}_4)E_{p_1}...E_{p_4} \frac{e^{-it(E_{k_1}+E_{k_2})}\,e^{+it'(E_{k_3}+E_{k_4})}}{E_{k_1}+E_{k_2}+E_{k_3}+E_{k_4}}\nonumber \\
 &&\int {\mathcal{D}}\tilde{\phi}\, \tilde{\phi}(\vec{p}_3)\tilde{\phi}(\vec{p}_4)\tilde{\phi}(\vec{k}_1)....\tilde{\phi}(\vec{k}_4)\tilde{\phi}(\vec{p}_1)\tilde{\phi}(\vec{p}_2) \Psi^{*(0)}_{0}[\tilde{\phi}]\Psi^{(0)}_{0}[\tilde{\phi}]\nonumber \\
 &=& -\lambda \, (2\pi)^3\delta^3(\vec{p}_1+\vec{p}_2+\vec{p}_3+\vec{p}_4) \,\,\frac{e^{-it(E_{p_1}+E_{p_2})}\,e^{+it'(E_{p_3}+E_{p_4})}}{E_{p_1}+E_{p_2}+E_{p_3}+E_{p_4}}
\end{eqnarray}
where $E_{p_i} = \omega_{p_i} = \sqrt{\vec{p}_i^2+m_i^2}$. In the first line of the above equation, some numerical constants and the normalization factors associated with the wavefunctional are suppressed.
For the S-matrix, one takes $t'\to\infty$ and  $t\to -\infty$ which can be implemented by taking $T = t' = -t$ and letting $T\to\infty$. Now use the identity
\begin{equation}
 \lim_{T \to \infty} \frac{e^{iET}}{E} = 2\pi i\,\delta(E),
\end{equation}
to obtain the four momentum delta function for the Scattering matrix element.

The important feature, relevant to us here, is the automatic appearance of the total energy ($E_T = E_{p_1}+E_{p_2}+E_{p_3}+E_{p_4}$) pole in the above computation. Instead of sending $T\to\infty$, if say we set $T = 0$, the time at which measurements are made, we get the desired total energy pole structure of the In-In correlation function. While what has been calculated is still the Scattering matrix element, the last but one step has the total energy pole quite evident. This is perhaps suggestive of the fact that Schrodinger functional representation and computations employing it would likely be a bridge between the amplitude methods which crucially rely on the wavefunction (and wavefunction coefficients). A rough comparison with the simplest of these calculations does provide a confirmation of this expectation. The task then would be to write down a Hamiltonian with the effective operators and compute the correlation functions in this wavefunctional approach.

\section{Conclusions and Discussion}
Effective Field Theories (EFTs) are used frequently as they enable separation of scales, and thus provide a clear statement of decoupling of physics at widely separated scales. While these issues are reasonably well understood for the EFTs for scattering and decay processes i.e. for In-Out amplitudes, the effective description for the In-In correlation functions has attracted less attention. Specifically, less attention has been paid to general aspects of such EFTs though there is no reason to expect that an effective description in such cases should not exist. In this work,
an attempt has been made to write down EFT for In-In correlation functions in the manner and spirit of the EFTs for In-Out amplitudes. Further, we've explicitly rederived the result that In-In and In-Out correlation functions are essentially the same, and then making use of this result, a matching calculation is presented in order to compute the Wilson coefficients of the EFT.  The most important feature now is the presence of boundary terms and these are included by treating the time and space derivatives differently even in a seemingly Lorentz invariant/covariant theory. The reason is very clear: the evaluation of the expectation value of the product of operators at specific time (often taken to be zero), breaks time translations while the spatial rotations are preserved. However, it still seems a bit odd that in the absence of the heavy field, the low energy theory should have been Lorentz invariant to start with. The measurements would have led to the effects of time translation breaking. At this point, the two varities of EFTs go different ways: in the In-Out case, the quantity of interest is the scattering amplitude which by itself is a Lorentz invariant object, and therefore, one is demanding that the full thery and the EFT yield the same Lorentz invariant quantity. Thus, the EFT also needs to be Lorentz invariant. In the In-In case, this is not so since the desired correlation function is not a Lorentz invariant object, and therefore, demanding that the EFT description gives the same result for the correlation function as the full theory necessitates allowing for Lorentz symmetry violation via the time translation breaking effects. This does have an impact on the operator basis, and we've seen with the explicit matching calculation that such operators are indeed needed in the EFT. One can explicitly check with a simple example like the one we've considered that allowing for Lorentz invariance breaking terms in an EFT for In-Out calculations yields Wilson coefficients which take the form and values such that the space and time components combine nicely into  Lorentz invariant final result. Without appealing to any particular type of interactions between the heavy and light fields, operators in terms of the light fields can be written directly, and a matching calculation can be performed against a given full theory. This was shown with a specific example at the tree level. The In-In EFT will have several new terms on top of the In-Out ones and more importantly, the coefficients of the time and spatial derivatives will be now different owing to the breaking of time translations, and thus Lorentz symmetry. It is also pointed out with a concrete example that Schrodinger picture captures the essence and is perhaps not unexpected as the wavefunction approach that is being pursued is closely related to it. Importantly, it captures the presence of the total energy pole. The construction here has been at the tree level and thus the immediate step forward would be explicitly check the construction at the loop level. The equilvalence between In-In and In-Out correlation functions will be put to a tighter test as now the issue of carefully handling vacuum bubbles will show up. Such an exercise will also lead to a better understanding of features inherent to the convetional way of computing In-In correlation functions, say via the Schwinger-Keldysh formalism. An important step even before that would be to identify redundant operators, possibly via field redefinitions, but explicitly keeping track of issues related to boundary terms. Application to cosmological correlators, at least at the tree level is straightforward with the only difficulty being in evaluating more complicated integrals. What would be very useful would be to rather extend this to finite temperature studies. The In-In correlation functions are usually computed using Schwinger-Keldysh method, and in the path integral evaluation one then has doubling of the fields: one set for the forward and backward path each. However, it is known that there are subtle issues if one wants to evaluate the expectation value of say a product of two fields/operators at the same time in the path integral formulation. This is another issue that needs to be properly addressed and understood for a clear and effective use of EFT techniques without leading to ambiguities. Several interesting directions remain to be explored as discussed above, not to mention that the notion of renormalization group evolution of the Wilson coefficients should also work out, albeit will need more care.

To summarise, while In-In correlation functions are encountered routinely in several areas, a direct calculation say via Schwinger-Keldysh formalism is employed and a large mass expansion is performed to get to effects pertaining to low energy theory. In this article, it is shown how to construct EFTs for In-In correltors directly based on the symmetries. Further, the computations can proceed the In-Out way, thanks to the equivalence between the two, hopefully simplifying computations in complex situations.\\

{\bf Acknowledgements}\\
This work is supported by the Department of Space (DOS), Government of India. Partial support under the MATRICS project (MTR/023/000442) from the Science $\&$ Engineering Research Board (SERB), Department of Science and Technology (DST), Government of India is acknowledged.

	\end{document}